\documentclass[runningheads]{llncs}
\usepackage[T1]{fontenc}
\usepackage{pifont}
\usepackage{oplotsymbl}
\usepackage{tabularx}
\usepackage{booktabs}
\usepackage{xltabular}
\usepackage{hyperref}
\hypersetup{breaklinks=true}
\usepackage{graphicx}
\usepackage{bbding}

\usepackage{color}

\renewcommand*{\orcidID}[1]{\href{https://orcid.org/#1}{\textsuperscript{\includegraphics[height=2.5ex]{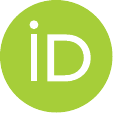}}}}
\begin{document}
\title{A Comparison of Kubernetes Compliance Standards and Configuration Scanners}
%
\titlerunning{Kubernetes Compliance Standards and Scanners}
%
\author{Michael Krieger\inst{1}\orcidID{0000-0003-3938-6492} \and
Markus Gierlinger\inst{2}\orcidID{0009-0004-9053-8599} \and
Farooq Shaikh\inst{1}\orcidID{0009-0005-3726-5373} \and
Mario Kahlhofer\inst{1}\orcidID{0000-0002-6820-4953}}
\authorrunning{M. Krieger et al.}
%
\institute{Dynatrace Research, Linz, Austria
\email{\{michael.krieger, farooq.shaikh, mario.kahlhofer\}@dynatrace.com}\\
\url{https://research.dynatrace.com} \and
Independent Researcher, Linz, Austria\\
\email{ma.gierlinger@gmail.com}}
%

\maketitle              
\begin{abstract}
Kubernetes has become the industry standard for orchestrating containers in microservice-based software architectures. While several hardening guidelines and scanning tools for securing Kubernetes clusters and deployments have emerged in recent years, their differing guidance and outputs often lead to inconsistent configuration and prioritization decisions. This work presents a systematic comparison of eight commonly used Kubernetes hardening guidelines. Through this comparison and the inclusion of best practices, we established a benchmark of 79 Kubernetes configuration recommendations and conducted the a structured empirical evaluation of ten popular static configuration scanning tools and their scoring outputs. Our findings reveal substantial disparities in the coverage of configuration issues across hardening guidelines and scanners, as well as inconsistencies in how configuration issues are scored and ranked by different scanners. These results highlight the need for more standardized, transparent, and consistent approaches to risk and severity assessment of Kubernetes configuration issues.

\keywords{Kubernetes \and Security Configuration \and Hardening Guidelines \and Static Analysis \and Configuration Scanning}
\end{abstract}
\section{Introduction}
\label{sec_introduction}
Kubernetes (k8s) is an open-source platform for automating the deployment, scaling, and management of containerized applications. Its resiliency and flexibility have driven widespread adoption across diverse domains, including 5G~networks~\cite{Carrión_2022,Ruiz_2023} and Internet of Things (IoT) environments such as software-defined vehicles~\cite{Shamim_2023}.  Its greatest impact, however, has been in cloud environments, where it has become the de~facto standard for service deployment.

In Kubernetes, application workloads are defined declaratively using manifests that describe a desired system state, which Kubernetes continuously reconciles with the actual cluster state. While this model integrates well with automation and simplifies operations, it requires substantial expertise. Developers must understand numerous configuration options, their defaults, and their security implications. Consequently, secure system configuration remains challenging beyond writing secure code~\cite{Dietrich_Krombholz_Borgolte_Fiebig_2018}, and misconfigurations often surface only after deployment in complex environments.

These challenges have led to the emergence of multiple Kubernetes hardening guidelines and standards, including the CIS Kubernetes Benchmark~\cite{CIS_2025}, the NSA/CISA Kubernetes Hardening Guide~\cite{NSA/CISA_2022}, and the Kubernetes Security Technical Implementation Guide (STIG)~\cite{DISA_2025}. Developed by different organizations, these guidelines vary in scope, level of detail, update frequency, and focus, ranging from high-level best practices to detailed configuration recommendations. The challenges posed by this fragmented landscape manifest in two ways. First, practitioners who rely on a single hardening guideline or scanner may develop a false sense of security, unaware of the significant coverage gaps that individual guidelines and tools exhibit. Second, achieving broad coverage requires combining multiple guidelines and tools, which substantially increases the operational workload. Depending on regulatory and organizational requirements, practitioners may further need to satisfy additional standards, compounding this effort.

Prior work has studied Kubernetes security practices and scanner performance~\cite{Shamim_Bhuiyan_Rahman_2020,Rahman_Shamim_Bose_Pandita_2023,Bose_Rahman_Shamim_2021,Kapetanidou_Nizamis_Votis_2025}, but has neither systematically compared hardening guidelines nor benchmarked scanner coverage against them. Moreover, while vulnerability scoring has been extensively researched~\cite{Allodi_Massacci_2014,Milousi_Kiriakidis_Mengidis_Rizos_Mazi_Voulgaridis_Votis_Tzovaras_2024,Koscinski_Nelson_Okutan_Falso_Mirakhorli_2025}, risk assessment approaches for Kubernetes configuration issues remain largely unexplored. To address these gaps, this paper presents a comparative analysis of Kubernetes hardening guidelines and benchmarks static configuration scanners with respect to coverage, detection accuracy, and risk assessment.

We investigate the following \emph{research questions}:
\begin{itemize}
    \item \textbf{RQ1:} \emph{Do different hardening guidelines and benchmarks provide the same configuration recommendations for Kubernetes?}
    \item \textbf{RQ2:} \emph{How accurately do different scanners detect common configuration issues in Kubernetes workloads?}
    \item \textbf{RQ3:} \emph{Do different scanners provide comparable risk assessments for common configuration issues?}
\end{itemize}

The primary contributions of our work are:

\begin{itemize}
    \item \textbf{Structured comparison of Kubernetes hardening guidelines}: We present a systematic comparison of hardening guidelines published by prominent organizations, identifying duplications, inconsistencies, and contradictions within and across guidelines.
    \item \textbf{Comprehensive empirical evaluation framework}: We introduce a data-driven framework for evaluating scanner accuracy, coverage, and severity scoring with respect to Kubernetes hardening guidelines.
    \item \textbf{Reusable benchmark manifests}: We develop benchmark manifests that enable reproducible evaluation of scanner accuracy, coverage, and severity scoring.
    \item \textbf{Actionable insights for practitioners}: We highlight substantial disparities in scanner coverage and scoring behavior, providing guidance for practitioners selecting tools to identify configuration issues in Kubernetes deployments.
\end{itemize}

\section{Kubernetes Architecture, Hardening Guidelines, and Scanners}
This section provides an overview of the architecture of Kubernetes and a description of selected hardening guidelines and scanners.

\subsection{Kubernetes}
Kubernetes is an open-source platform for automating the deployment, scaling, and management of containerized applications. Application workloads and supporting resources are defined declaratively using manifests, which describe a desired state that Kubernetes continuously reconciles with the actual cluster state.

\begin{figure}[tb]
  \centering
  \includegraphics[width=0.8\columnwidth]{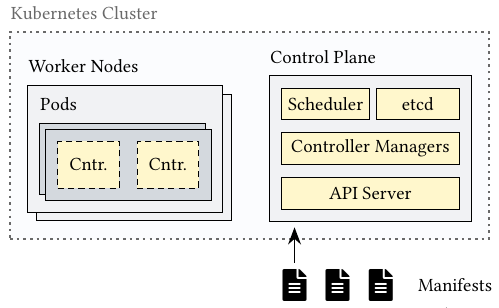}
  \caption{Typical components of a Kubernetes cluster.}
  \label{fig:kubernetes}
\end{figure}

Architecturally, a Kubernetes cluster is divided into a control plane and a data plane. The control plane provides orchestration and management functionality through components such as the API server, controller managers, scheduler, and etcd. It enforces authentication and authorization (typically via role-based access control) and uses controllers to reconcile the desired state defined in manifests with the cluster’s current state.

The data plane hosts application workloads on worker nodes. Containers run within pods, the smallest deployable units in Kubernetes, while higher-level resources such as deployments and services manage replication, fault tolerance, load balancing, and stable network endpoints. Each worker node runs a kubelet agent that manages the lifecycle of workloads assigned to that node~\cite{Martin_Hausenblas_2021}.

\subsection{Compliance and Security Guidelines}
The Kubernetes Policy Working Group emphasizes the importance of adhering to both internal and external regulatory standards, as well as security best practices, particularly for Kubernetes workloads that process sensitive data or perform critical functions~\cite{Ramanathan_2023}. Because data and computing resources in Kubernetes environments are often globally distributed, achieving regulatory compliance is especially challenging, as requirements frequently depend on the specific jurisdictions in which data is processed.

\begin{figure}[tb]
  \centering
  \includegraphics[width=0.8\columnwidth]{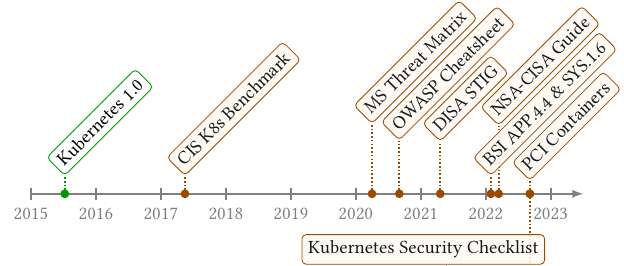}
  \caption{Timeline of first releases of K8s security standards.}
  \label{fig:standards-timeline}
\end{figure}

For our study, the selection of standards followed a systematic process. First, we examined the security scanners analyzed in Section~\ref{sec:scanners} to identify the regulatory frameworks they support. We then reviewed widely recognized regulatory standards, including PCI DSS, HIPAA, GDPR, and SOC~2, together with their associated technical guidance. In addition, we considered recommendations issued by authoritative institutions in the United States and the European Union, such as the NSA, NIST, ENISA, and the PCI Security Standards Council (PCI~SSC). Vendor-specific guidelines were excluded to avoid product-specific bias. This study does not aim to provide an exhaustive survey of all existing frameworks or guidelines; instead, we focus on widely adopted, reputable, and influential sources used by practitioners. The following section provides an overview of the eight security standards selected for analysis~(Figure~\ref{fig:standards-timeline}). Additional sources reviewed (e.g., ENISA container security guidance, Docker CIS Benchmark) either fell outside Kubernetes scope or repeated recommendations already captured, suggesting thematic saturation.


\begin{enumerate}
    \item \textbf{CIS Kubernetes Benchmark:} Since 2017, the Center for Internet Security~(CIS) has maintained a dedicated benchmark for Kubernetes, covering Kubernetes distributions and managed Kubernetes services. The CIS Kubernetes Benchmark is regularly updated and currently includes 131 recommendations (version~1.12). 
    \item \textbf{Microsoft Threat Matrix for Kubernetes:} In 2020, Microsoft released the Threat Matrix for Kubernetes~\cite{ms_threat_matrix_for_kubernetes_2021}, which builds on the MITRE ATT\&CK framework and identifies 35 mitigation actions mapped to 40 attack techniques. The threat matrix does not assign explicit risk or severity levels to the identified threats or mitigations.
    \item \textbf{OWASP Kubernetes Security Cheat Sheet:} In 2020, the Open Worldwide Application Security Project (OWASP) published the initial version of the Kubernetes Security Cheat Sheet~\cite{owasp_kubernetes_cheat_sheet_2025}, which has since been updated regularly. Eight sections cover the full lifecycle of a Kubernetes deployment.
    \item \textbf{DISA Kubernetes STIG:} Since 2021, the Defense Information Systems Agency (DISA) maintains a dedicated Security Technical Implementation Guide (STIG) for Kubernetes~\cite{DISA_2025}, which is updated regularly. Each of its 91 configuration rules includes a description, rationale, audit procedures, and remediation actions and an assignment to a severity category (Category I to III), corresponding to high, medium, and low severity.
    \item \textbf{NSA-CISA Kubernetes Hardening Guideline:} In 2022, the National Security Agency (NSA) and the Cybersecurity and Infrastructure Security Agency (CISA) jointly published the Kubernetes Hardening Guide~\cite{NSA/CISA_2022}. While an appendix provides concrete configuration examples, the main body of the guide~--- comprising 34 sections~--- focuses on general hardening strategies rather than prescriptive configuration rules.
    \item \textbf{BSI APP.4.4 and SYS.1.6:} In 2022, the German Federal Office for Information Security (BSI) first published Kubernetes-related guidance as part of its IT-Grundschutz Compendium in sections \emph{APP.4.4~Ku\-ber\-ne\-tes}~\cite{BSI_APP_4_4_2023}, comprising 21 requirements, and \emph{SYS.1.6 Con\-tai\-ne\-ri\-sie\-rung}~\cite{BSI_SYS_1_6_2023}, comprising 26 requirements and categorized into basic, standard, and increased protection levels.
    \item \textbf{PCI Guidance for Containers and Container Orchestration Tools:} In 2022, the PCI Security Standards Council (PCI~SSC) released a supplement providing best practices for containers and container orchestration tools~\cite{PCI_SIG_2022}. The document identifies 42 potential threats and proposes 49 technology-agnostic best practices to mitigate these threats.
    \item \textbf{Kubernetes Security Checklist:} The Kubernetes project maintains several security hardening resources, including a general \emph{security checklist}~\cite{Kubernetes_security_checklist_2025} and an \emph{application security checklist}~\cite{Kubernetes_application_security_checklist_2025}. Together, these checklists comprise 76 recommendations, ranging from high-level guidance to specific configuration settings, including references to 12 recommended admission controllers.
\end{enumerate}

\subsection{Scanners}
\label{sec:scanners}
Many recommendations from the hardening guidelines described above can be verified automatically. This work focuses on tools for static analysis of Kubernetes manifests, which can be integrated into the development and pre-deployment phases of Kubernetes clusters and applications.

To identify relevant tools, we conducted a comprehensive search across multiple sources, including the CNCF Landscape~\cite{CNCFLandscape}, GitHub repositories, and online materials such as blog posts and industry reports. We used targeted search terms—including ``Kubernetes'', ``k8s'', ``security'', ``misconfiguration'', ``compliance'', and ``scanner''~--- to ensure broad coverage. In addition, we reviewed industry analyses such as the Gartner Application Security Testing Magic Quadrant~\cite{Gartner25Ast}, the Forrester Wave Software Composition Analysis~\cite{Forrester24Sca}, and the Forrester Wave Static Application Security Testing~\cite{Forrester25Ast} to identify prominent tools and industry trends. We limited our scope to open-source tools, including open-source offerings maintained by commercial vendors, as these tools are freely available to the Kubernetes community and often incorporate guidance derived from established security standards. We also added commercial tools with open-source CLI wrappers to our scope, but excluded purely proprietary and paid solutions. We acknowledge that this scope restriction means our findings are specific to the open-source scanner ecosystem; commercial scanners may exhibit different coverage and scoring behavior, which warrants separate investigation. Tools requiring a running cluster (e.g., kube-bench, KubiScan) were also excluded. Table~\ref{tab:tools} provides an overview of the 10 scanners analyzed in this study. The selected tools are listed below:


\begin{figure}[tb]
  \centering
  \includegraphics[width=0.8\columnwidth]{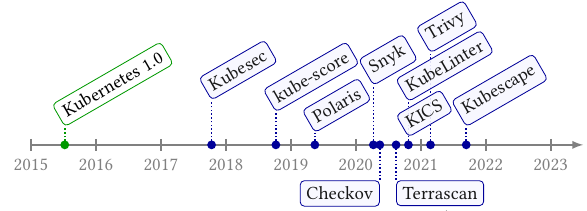}
  \caption{Timeline of first releases of K8s security scanners.}
  \label{fig:scanners-timeline}
\end{figure}

\begin{enumerate}
    \item \textbf{Kubesec}~\cite{kubesec_2025} validates Kubernetes manifests using \emph{kubeconform}~\cite{kubeconform_2025} and can be run as a standalone tool.
    
    \item \textbf{kube-score}~\cite{kubescore_2025} is a community-driven project with a single person serving as publisher and main contributor, distinguishing it from corporate-backed alternatives.
    
    \item \textbf{Polaris}~\cite{polaris_2025} functions as a policy engine that can be deployed as a dashboard, admission controller, or standalone scanner, with rules focusing on pod configuration.
    
    \item \textbf{Snyk}~\cite{snyk_2025} is a broader (commercial) security platform where Kubernetes scanning is part of its Infrastructure-as-Code (IaC) capabilities, which also requires connectivity to Snyk's platform. Snyk can scan Kubernetes manifests and Helm charts.
    
    \item \textbf{Checkov}~\cite{checkov_2025} is an IaC and software composition analysis (SCA) tool from Palo Alto. Checkov supports scanning different technologies, including Kubernetes manifests and Helm charts.
    
    \item \textbf{Terrascan}~\cite{terrascan_2025} has been available as an archived project since November 2025, meaning it no longer receives active development. Terrascan scans manifests and Helm charts.
    
    \item \textbf{KICS}~\cite{kics_2025} (\emph{Keeping Infrastructure as Code Secure}) operates fully offline without requiring connectivity to external services and can scan multiple different technologies.
    
    \item \textbf{KubeLinter}~\cite{kubelinter_2025}, sponsored by Red Hat and maintained by the StackRox community, can be run as a standalone tool on manifests and Helm charts. It provides no severity ratings.
    
    \item \textbf{Trivy}~\cite{trivy_2025} combines multiple scanning capabilities for manifests and Helm charts, including detecting vulnerable dependencies, sensitive data, secrets, and IaC misconfigurations.
    
    \item \textbf{Kubescape}~\cite{kubescape_2025}, originally developed by ARMO, is a CNCF incubating project that can run as standalone IaC scanner but also be used as admission controller or vulnerability scanner. Its severity scores are similar to CCSS~\cite{36061} and CVSS~\cite{FIRST_CVSS_v4}.
\end{enumerate}

\begin{table}
\centering
\caption{Overview of Kubernetes Security Scanning Tools}
\label{tab:tools}
\newcommand{\xmark}{{\color{black!10}$\circlet$}}
\newcommand{\tmark}{{\color{black!65}{$\circletfillhl$}}}
\newcommand{\cmark}{$\circletfill$}
\begin{tabularx}{1\textwidth}{@{\extracolsep{\fill} }lclclr}
\toprule
\textbf{Name} & \textbf{Year} & \textbf{Maintainer} & \textbf{OSS} & \textbf{Severity Ratings} & \textbf{Rules} \\
\midrule
Kubesec~\cite{kubesec_2025} & 2017 & ControlPlane & \cmark &  -30 to 1 & 20+ \\
&&&&(critical to low)&\\
kube-score~\cite{kubescore_2025} & 2018 & Community & \cmark &  1/5/7/10 & 30+ \\
&&&&(critical/warning/low/ok)&\\
Polaris~\cite{polaris_2025} & 2019 & Fairwinds & \cmark & danger/warning & 40+ \\
Snyk~\cite{snyk_2025} & 2020 & Snyk & \tmark & low/medium/high/critical & 40+ \\
Checkov~\cite{checkov_2025} & 2020 & Palo Alto & \tmark & low/medium/high & 110+ \\
Terrascan~\cite{terrascan_2025} & 2020 & Tenable & \cmark & low/medium/high & 30+ \\
KICS~\cite{kics_2025} & 2020 & Checkmarx & \cmark & info/low/medium/high & 140+ \\
KubeLinter~\cite{kubelinter_2025} & 2020 & StackRox & \cmark & --- & 60+ \\
Trivy~\cite{trivy_2025} & 2021 & Aqua Security & \cmark & low/medium/high/critical & 160+ \\
Kubescape~\cite{kubescape_2025} & 2021 & Armo / CNCF & \cmark & 0 to 10 & 260+ \\
&&&&(low to critical)&\\
\bottomrule
\end{tabularx}
\end{table}

\section{Experiment Setup}
This section describes the experiment design: how we categorized recommendations (Section~\ref{subsec_unique_recommendations}), defined benchmark manifests (Section~\ref{subsec_benchmark_manifests}), derived severity scores (Section~\ref{subsec_scoring_method}), and analyzed results (Section~\ref{subsec_data_analysis}).

\subsection{Identification of unique configuration recommendations}
\label{subsec_unique_recommendations}
To address \emph{RQ1 (consistency across hardening guidelines)}, we identified recommendations recurring across multiple hardening guidelines. While the Common Configuration Enumeration~(CCE)~\cite{nist-cce} provides a standardized mechanism for uniquely identifying configuration recommendations, and NIST maintains a registry of assigned identifiers\footnote{\url{https://ncp.nist.gov/cce}}, no CCEs are currently defined for Kubernetes configuration recommendations. Moreover, only a small subset of guidelines assigns unique identifiers to Kubernetes recommendations (e.g., STIG). We therefore manually compared all selected guidelines, checked the individual recommendations for equivalence, and compiled a unified set of unique recommendations. For guidelines providing explicit configuration checks, we validated equivalence based on: (i) the configuration parameter described, and (ii) the recommended value or check procedure. For guidelines providing higher-level recommendations, we assessed whether our extracted configuration recommendations were covered by these statements. Where guidelines provided unique identifiers, we used these for cross-referencing.

Given our focus on cluster and workload configuration, we aligned only recommendations focusing on Kubernetes architectural components. We adopted the categorization scheme of the CIS Kubernetes Benchmark, distinguishing recommendations related to the control plane, worker nodes, and policies. Recommendations not directly attributable to a specific component but still relevant to Kubernetes were classified as \emph{General Policies}. Recommendations outside the scope of Kubernetes cluster and workload configuration~--- such as those related to CI/CD pipelines, image registries, or third-party tools~--- were excluded.

This categorization enables a structured comparison of hardening guidelines by (i) separating control-plane and data-plane recommendations, (ii) identifying thematic emphases across guidelines, and (iii) consolidating configuration recommendations across multiple sources.

\subsection{Definition of Benchmark Manifests}
\label{subsec_benchmark_manifests}
Our benchmark focuses on configuration issues expressible in Kubernetes manifest files for data-plane resources, including aspects of role-based access control (RBAC), service accounts, deployments, and network policies. We acknowledge this as a limitation. Future work should address control-plane security through complementary runtime analysis approaches.

For generating benchmark manifests we started with 67 configuration recommendations for data-plane resources extracted from hardening guidelines. We added 12 additional security-relevant recommendations derived from operational best practices identified through the authors' combined experience in Kubernetes security engineering. These additional recommendations were validated through the same expert review process described in Section~\ref{subsec_scoring_method} and are documented in the benchmark repository, resulting in a total of 79 configuration recommendations. For each recommendation, we created manifests violating exactly one configuration rule. When multiple violation variants existed for a single recommendation, we encoded each variant in a separate manifest. This process resulted in a total of 241 manifests used for scanner benchmarking. All manifests were created using the AWS Cloud Development Kit for Kubernetes (cdk8s)~\cite{cdk8s_2025} with Python bindings, which is a simple wrapper facilitating syntactically correct manifest generation. We first created secure default configurations~(non-root, disabled host namespaces, read-only root filesystem, explicit resource limits) and then introduced a single deliberate violation per manifest by modifying the relevant parameter to a less secure setting. Each manifest was annotated with metadata, including (i) references to the corresponding hardening guideline recommendations, (ii) the exact manifest paths containing the configuration issue, (iii) an expectation of whether a scanner should report the issue, (iv) an expert-assigned severity category, and (v) an expert-assigned severity score (see Section~\ref{subsec_scoring_method}).

To enable consistent evaluation across all scanners, we developed an application that orchestrates test executions and analyzes scanner outputs. The application includes a wrapper for each scanner to extract metadata such as version information, supported execution modes, and output formats. Because several scanners produced multiple findings or lacked sufficient information to distinguish true from false positives, we incorporated scanner-specific metadata into the evaluation process. For example, when a scanner did not report precise manifest locations for findings, we manually mapped scanner rules to the corresponding manifest paths. We used JSON to represent these mappings in a machine-readable format and added appropriate post-processing to the scanner wrappers. This mapping and post-processing are implemented as part of a unified evaluation pipeline. All scanners were evaluated using their default configurations and built-in rule sets without custom policies or security overrides.

\subsection{Scoring Configuration Issues}
\label{subsec_scoring_method}
We started with a literature review of configuration scoring systems. The most relevant contribution is the work on the Common Configuration Scoring System (CCSS) by Scarfone and Mell~\cite{Scarfone_Mell_2008} published by the National Institute of Standards and Technology (NIST) in 2010~\cite{36061}. CCSS was designed to complement established vulnerability scoring systems such as the Common Vulnerability Scoring System (CVSS) and the Common Misuse Scoring System (CMSS)~\cite{LeMay_Scarfone_Mell_2012}.

For this study, we applied CCSS to the identified configuration recommendations. As no official CCSS scores are available for Kubernetes configurations, we engaged three domain experts with demonstrable hands-on experience in Kubernetes security engineering to independently assess each configuration recommendation. Following individual assessments, the experts resolved disagreements through a majority-vote consensus process to derive a final expert-curated CCSS-based severity score for each recommendation. We acknowledge that CCSS has not been revised since its initial publication in 2010; however, it remains the only standardized scoring framework designed specifically for configuration issues, and we use it here as a structured comparative baseline rather than as an authoritative ground truth.

To enable comparison with scanner-reported severities, we normalized the severity outputs of all scanners. This normalization involved (i) identifying the severity categories used by each scanner, (ii) mapping them to the qualitative severity scale defined by CVSS, and (iii) translating these qualitative ratings into representative values within CVSS severity ranges. The resulting normalized scanner scores and the expert-curated CCSS scores were then used in the experimental analysis to assess the consistency of severity assessments across scanners and their alignment with expert judgment.

\subsection{Data Analysis}
\label{subsec_data_analysis}
\paragraph{RQ1 (consistency across hardening guidelines):} We mapped recommendations across guidelines following the methodology described in Section~3.1.

\paragraph{RQ2 (detection quality of different scanners):}
\label{sec:method_rq2} To answer this question, we scanned our curated set of benchmark manifests using the selected scanners. We identified true positives, false positives, and false negatives by comparing the YAML path of each injected configuration issue in the benchmark manifests with the YAML path referenced by the scanner findings. From the resulting TP/FP/FN counts, we computed precision (fraction of alerts that are true positives) and recall (fraction of benchmark issues detected) for each tool, along with two aggregate metrics:
\begin{itemize}
\item \emph{Coverage}, defined as the fraction of benchmark configuration issues detected by a tool (i.e., the overlap between detected issues and the full benchmark set). Low coverage indicates that a tool is specialized in specific categories or implements only a limited set of checks.
\item \emph{F1-score}, defined as the harmonic mean of precision and recall, summarizing detection performance in a single metric that weights false positives and false negatives equally.
\end{itemize}

\paragraph{RQ3 (alignment of severity scoring):} 
\label{sec:method_rq3} To address this question, we built on the benchmark used for RQ2. We included only results where the scanner provided a severity assessment. For scanners reporting qualitative severity categories, we mapped categories to numeric values as follows: \emph{critical} to a score of 9.5, \emph{high} or \emph{danger} to 8.0, \emph{medium} or \emph{warning} to 6.0, \emph{low} to 2.0, \emph{info} or \emph{ok} to 0.1 and \emph{skip} to 0.0, following the CVSS guideline.

For each finding, we computed the similarity ($s$) between a scanner severity score ($c_s$) and the expert score ($c_e$) using:  

\begin{equation}
  s = 1 - \frac{abs(c_s - c_e)}{10}
\end{equation}

The resulting similarity score ranges from 0 (maximally different) to 1 (equal). For each scanner, we then computed the mean and variance of similarity scores to quantify how consistently the scanner’s severity assessments align with expert scores. To identify potential scoring bias, we also computed the mean signed deviation (MSD) between scanner scores and expert scores. MSD captures whether a scanner’s severity assessments are, on average, higher or lower than the CCSS-based scores curated by the expert group.

\begin{table}
  \centering
  \caption{Overview of the analyzed hardening guidelines, number of recommendations or sections aligned to Kubernetes' architectural components and concepts, identified unique recommendations across all analyzed hardening guidelines, and total number of recommendations/section which were mapped to architectural components and concepts.}
  \label{tab:guidelines}
  \setlength{\tabcolsep}{3pt}
  \begin{tabularx}{\textwidth}{Xlrrrrrrrr|r}
    \toprule
     & & \rotatebox{90}{\bfseries CIS 1.12} & \rotatebox{90}{\bfseries NSA-CISA} & \rotatebox{90}{\bfseries STIG} & \rotatebox{90}{\bfseries BSI} & \rotatebox{90}{\bfseries PCI} & \rotatebox{90}{\bfseries MS TM.} & \rotatebox{90}{\bfseries K8s Chklst.} & \rotatebox{90}{\bfseries OWASP} & \rotatebox{90}{\textbf{extracted}}\\
    \midrule
    & Master Node & 21 & 2 & 12 & 1 & 1 & 1 & -- & 1 & \textbf{24}\\
    & API Server & 30 & 7 & 26 & 3 & 5 & 6 & 15 & 4 & \textbf{47}\\
    & Controller Manager & 7 & 3 & 6 & 1 & 6 & 1 & 1 & 1 & \textbf{9}\\
    Control & Scheduler & 2 & 2 & 3 & -- & 2 & 1 & -- & 1 & \textbf{4}\\
    & etcd & 7 & 1 & 9 & 1 & 3 & 1 & 3 & 1 & \textbf{8}\\
    & AuthN \& AuthZ & 3 & 1 & -- & 1 & 2 & 1 & -- & 1 & \textbf{3}\\
    & Logging & 2 & 2 & 2 & -- & 2 & 1 & 1 & 3 & \textbf{4}\\
    \cmidrule(lr){1-11}
    & Worker Node & 10 & 2 & 15 & 1 & 1 & -- & -- & -- & \textbf{15}\\
    Worker & Kubelet & 14 & 4 & 11 & 2 & 6 & -- & 1 & 3 & \textbf{18}\\
    & kube-proxy & 1 & -- & -- & -- & 1 & -- & -- & -- & \textbf{1}\\
    \cmidrule(lr){1-11}
    & RBAC & 13 & 4 & -- & 3 & 6 & 3 & 10 & 2 & \textbf{18}\\
    & Pod Security & 12 & 5 & 3 & 4 & 5 & 5 & 15 & 5 & \textbf{17}\\
    Policies & Network Policies & 2 & 3 & -- & 2 & 1 & 3 & 6 & 2 & \textbf{7}\\
    & Secrets Management & 2 & -- & 1 & 1 & 3 & 2 & 3 & 2 & \textbf{3}\\
    & Ext. Adm. Control & 1 & -- & -- & 1 & 2 & 1 & 2 & 1 & \textbf{2}\\
    & General Policies & 4 & 3 & 5 & 4 & 5 & 2 & 5 & 3 & \textbf{10}\\
    \specialrule{0.12em}{0.5em}{0.5em}
    \bfseries Rules & \bfseries Total & 131 & 34 & 91 & 47 & 49 & 35 & 76 & 53 & \textbf{190}\\
    & \bfseries Mapped & 131 & 23 & 88 & 15 & 31 & 22 & 60 & 24 & \\
    & \bfseries Mapped (in \%) & 100\% & 68\% & 97\% & 32\% & 63\% & 63\% & 79\% & 45\% & \\
    \bottomrule
  \end{tabularx}
\end{table}

\section{Evaluation of Guidelines and Scanners}

\subsection{RQ1: Evaluating consistency of recommendations}

Following the methodology outlined in Section~3.1, we mapped all extracted recommendations to Kubernetes architectural components (Table~\ref{tab:guidelines}), excluding peripheral aspects such as CI/CD pipelines and image registries. In contrast, only 32\% of the recommendations in \emph{BSI APP.4.4 Kubernetes} and 45\% of those in the \emph{OWASP Kubernetes Security Cheat Sheet} could be mapped to our categorization.

In total, we identified 190 distinct recommendations across the selected hardening guidelines. As shown in \emph{Table \ref{tab:guidelines}}, most recommendations target the API server (47), the master node (24), the kubelet (18), RBAC and the use of service accounts (18), pod security (17), and the worker node (15). The least covered component is kube-proxy, with only a single recommendation.

We did not identify any conflicting recommendations across guidelines published by different organizations. However, when examining individual guidelines, we identified contradictory statements within the Kubernetes Security Checklists published by the Kubernetes project. Specifically, the general \emph{security checklist} states that \emph{memory limits should be set for the workloads with a limit equal or inferior to the request}, whereas the \emph{application security checklist} states that \emph{Memory limit should be set for the workloads with a limit equal to or greater than the request}. In addition, analysis of the \emph{Kubernetes STIG v2R4} revealed several duplicate rules. For example, rules CNTR-K8-000860 and CNTR-K8-003110 both address ownership of files in the \texttt{/etc/kubernetes/manifests/} directory. Similarly, rules CNTR-K8-000880 and CNTR-K8-003240 both concern ownership of the kubelet configuration file, while rules CNTR-K8-000890 and CNTR-K8-003230 both address file permissions for the kubelet configuration file.

Overall, our analysis indicates that the CIS Kubernetes Benchmark~--- currently the most widely adopted hardening guideline~--- provides a solid baseline in the sense that it achieves the broadest categorical coverage across Kubernetes components and is most frequently referenced by the scanners evaluated in this study. However, more than 50 additional recommendations for hardening Kubernetes clusters and deployments can be derived from the other analyzed guidelines. We identified no contradictions across different guidelines and only three duplicates and one contradiction within single guidelines, indicating a generally high degree of consistency across existing hardening guidelines.

\subsection{RQ2: Detection quality and coverage of different scanners}


We evaluated a total of 10 scanners using the benchmark manifests. For the calculation of coverage and F1-score, alerts were treated as binary classifications. An alert was considered a true positive only if the YAML path reported by the scanner matched the YAML path of the injected configuration issue in the manifest. If multiple alerts were raised for a manifest, the result was considered a true positive if at least one alert matched the correct path. Alerts without a matching path were classified as false positives. If a recommendation was represented by multiple variant manifests, a scanner was credited with detecting the issue if it successfully flagged at least one of the variants. If a scanner failed to raise an alert for all manifest variants representing a recommendation, the result was classified as a false negative and, consequently, as a lack of coverage. 

As shown in \emph{Table~\ref{tab:scanner_comparison}}, only two scanners~--- \emph{Trivy} and \emph{KICS}~--- achieved an F1-score of 0.69 or higher, representing the strongest overall balance of precision and recall among all evaluated scanners, while also achieving coverage of at least 50\%. Both scanners include more than 140 Kubernetes-specific rules, suggesting a mature rule set. An outlier in this analysis is \emph{Polaris}, which achieved competitive F1-score and high coverage despite providing fewer than 50 rules. This result indicates that Polaris's rules are well-targeted at the configuration categories represented in the benchmark, with fewer spurious alerts compared to larger rule sets~--- suggesting that rule quality and specificity are more important than rule count for overall detection performance. All remaining scanners with fewer than 100 rules achieved coverage of 32\% or less. Conversely, \emph{Kubescape} illustrates that a larger rule set does not necessarily yield better results: despite providing the largest number of built-in rules, it did not achieve the highest coverage or F1-score.

Additionally, we grouped benchmark manifests by category. More than half of the scanners included checks for exposed secrets (via environment variables or ConfigMaps). RBAC and service account configurations~--- represented by 26 configuration recommendations~--- showed particularly low coverage, with no scanner exceeding 40\%. Network policy configurations were supported by many scanners, yet none correctly identified more than 50\% of the corresponding benchmark manifests. In the area of workload and deployment definitions, six scanners detected more than 60\% of configuration issues. No scanner detected missing pod-security labels in namespace definitions. 

\begin{table*}
  \centering
  \caption{List of scanners in alphabetical order with number of built-in rules for Kubernetes, calculated coverage and F1 scores, and details for different recommendation categories, with the number of manifests per category in parentheses.}
  \label{tab:scanner_comparison}
  \setlength{\tabcolsep}{3pt}
  \newcommand{\best}[1]{\textbf{\textcolor{orange!80!black}{#1}}}
  \begin{tabularx}{\textwidth}{Xrrr|rrrrrrr}
    \toprule
     \bfseries Scanner & \bfseries Built-in rules & \bfseries Coverage(\%) & \bfseries F1-Score & \rotatebox{90}{\bfseries RBAC (26)} & \rotatebox{90}{\bfseries Pod Security (33)} & \rotatebox{90}{\bfseries Network (5)} & \rotatebox{90}{\bfseries Secrets (2)} & \rotatebox{90}{\bfseries Admission (2)} & \rotatebox{90}{\bfseries General (11)} \\
    \midrule
    Checkov     & 113           & 38.0          & 0.550         & 5             & 18                & --             & --             & 1             & 6 \\
    KICS        & 142           & \best{54.4} & \best{0.705}& \best{12}   & 22                & --             & 1             & 1             & \best{7} \\
    kube-score  & 39            & 25.3          & 0.354         & --             & 11                & \best{2}    & 1             & \best{2}    & 4 \\
    KubeLinter  & 63            & 30.4          & 0.466         & 2             & 15                & 1             & 1             & 1             & 4 \\
    Kubescape   & \best{262}  & 43.0          & 0.481         & 3             & 20                & 1             & \best{2}    & \best{2}    & 6 \\
    Kubesec     & 22            & 24.1          & 0.388         & 2             & 15                & --             & --             & --             & 2 \\
    Polaris     & 44            & 51.9          & 0.589         & 7             & 23                & \best{2}    & \best{2}    & \best{2}    & 5 \\
    Snyk        & 45            & 27.8          & 0.420         & 5             & 15                & --             & --             & 1             & 1 \\
    Terrascan   & 35            & 31.6          & 0.481         & 1             & 16                & --             & --             & 1             & \best{7} \\
    Trivy       & 169           & 53.2          & 0.694         & 10            & \best{25}       & --             & 1             & 1             & 5 \\
    \bottomrule
  \end{tabularx}
\end{table*}

\subsection{RQ3: Alignment of severity scoring}

Nine of the ten evaluated scanners provided a risk or severity assessment for identified configuration issues. \emph{KubeLinter} did not provide a severity score, while \emph{Kubescape} was the only scanner to report numeric scores in the range of 0.0 to 10.0, consistent with the CCSS scale. All other scanners reported qualitative severity categories (e.g., critical, high, danger, medium, warning, low, info, ok, skip).

Across the benchmark, scanners provided severity assessments for 61 of the 79 configuration recommendations. Only one configuration issue received an assessment from all scanners, while 17 configuration issues were assessed by only a single scanner. After normalizing severity categories to CCSS scores, substantial discrepancies became evident. For 20 configuration issues, the difference between the lowest and highest assigned severity score was 5.0 or greater, and for nine issues the difference exceeded 7.0. The largest severity disagreements clustered around security context configurations (average range of 7.5 CCSS points) and resource limit definitions (average range of 5.2 points). For example, the missing readOnlyRootFilesystem specification in a container's security context received severity ratings ranging from \emph{info} (0.1, Kubesec) to \emph{critical} (9.5, kube-score)~--- a 9.4-point spread for the same configuration issue. In contrast, RBAC permission checks showed the highest cross-scanner agreement, with an average severity range of 2.6 CCSS points. Notably, security context checks were the worst-aligned category for four of nine scanners (Checkov, KICS, kube-score, Terrascan) yet the best for Snyk and Trivy, revealing fundamental disagreement on how to classify these issues. When mapped to CVSS v3/v4 qualitative levels (Critical, High, Medium, Low, None), these differences would span at least three severity levels for the same configuration issue.


We further compared scanner-assigned severity scores with the expert-curated CCSS scores, where the expert scores serve as a relative reference baseline. Four scanners produced, on average, higher severity ratings than the experts, while five scanners produced lower ratings (see Table~\ref{tab:scanner_scoring_comparison}). \emph{Terrascan} showed the closest overall alignment with expert ratings, with a mean signed deviation (MSD) of 0.086. The most stable scanners in terms of similarity to expert ratings were \emph{Kubescape} (similarity=0.857, MSD=0.786, variance=0.010), \emph{Polaris} (similarity=0.845, MSD=0.984, variance=0.013), and \emph{Terrascan} (similarity=0.830, MSD=0.086, variance=0.013). 

\begin{table}
  \centering
  \caption{List of scanners with number of findings, average similarity score, variance of similarity, and mean signed deviation (MSD) of similarity}
  \label{tab:scanner_scoring_comparison}
  \setlength{\tabcolsep}{6pt}
  \newcommand{\best}[1]{\textbf{\textcolor{orange!80!black}{#1}}}
  \begin{tabularx}{\columnwidth}{Xrrrr}
    \toprule
     \bfseries Scanner    & \bfseries Findings      & \bfseries Avg. Similarity        & \bfseries Sim. Variance          & \bfseries Sim. MSD\\
    \midrule
    Checkov     & 27            & 0.743             & 0.023             & -2.341 \\
    KICS        & 35            & 0.784             & 0.028             & -1.074 \\
    kube-score  & 13            & 0.624             & 0.033             & 2.515 \\
    Kubescape   & 25            & \best{0.857}    & \best{0.010}    & 0.786 \\
    Kubesec     & 17            & 0.653             & 0.059             & -2.647 \\
    Polaris     & 31            & 0.845             & 0.013             & 0.984 \\
    Snyk        & 19            & 0.802             & 0.018             & -1.521 \\
    Terrascan   & 21            & 0.830             & 0.013             & \best{0.086} \\
    Trivy       & \best{38}    & 0.784             & 0.018             & -0.700 \\
    \bottomrule
  \end{tabularx}
\end{table}

\section{Threats to Validity}
We follow well-established guidelines for experimentation in software engineering~\cite{Wohlin_Runeson_Höst_Ohlsson_Regnell_Wesslen_2024} and discuss four types of threats to validity: conclusion validity, internal validity, construct validity, and external validity. Internal validity does not apply to our work, as it concerns incorrect cause-effect relationships, which are not the focus of this study. The remaining threats apply as follows:
\begin{itemize}
\item \emph{Conclusion validity}: Our benchmark contains only configuration checks related to the data plane. Consequently, coverage does not extend to all possible Kubernetes configuration issues. While completeness cannot be demonstrated conclusively, the benchmark remains extensible to include control-plane configurations in future work. Furthermore, vertical-specific compliance regulations such as PCI DSS, HIPAA, and similar standards were intentionally excluded, as their scope is highly domain-specific and only indirectly related to the objectives of this study. Also, as our benchmark is derived from existing compliance frameworks and established tools, the results are inherently biased toward known and widely adopted controls.
\item \emph{Construct validity} concerns the appropriateness of conclusions drawn from the underlying measurements. In our case, the catalog of configuration recommendations is intended to support a more systematic implementation and evaluation of compliance benchmarks. Assessing whether this goal is achieved would require a separate empirical investigation, which is beyond the scope of this work and represents an avenue for future research.
\item \emph{External validity} relates to the generalizability of the findings. Our study is specific to Kubernetes and may not directly apply to other container management or orchestration technologies. Moreover, the benchmark encodes exactly one configuration issue per check, while providing multiple variants for certain issues, all of which are weighted equally in the analysis. This introduces an inherent bias in the resulting performance metrics, favoring categories with a larger number of checks or variants.
\end{itemize}

Beyond these categories, a general threat to validity arises from temporal currency. The benchmark represents a snapshot in time and must be continuously maintained to remain relevant, as Kubernetes and its ecosystem evolve rapidly. For example, at the beginning of this research, Kubernetes supported \emph{Pod Security Policies (PSP)} as a native admission control mechanism. During the course of this study, PSPs were deprecated and are no longer viable. Current best practices instead recommend the use of third-party admission controllers or, at a minimum, the enforcement of \emph{Pod Security Standards (PSS)}.

\section{Discussion}

The main findings of this study are the limited overlap among Kubernetes hardening guidelines, the fragmented coverage and detection accuracy of configuration scanners, and the lack of a coordinated approach to configuration risk assessment. This section summarizes these findings and derives implications for practitioners and future research.

\subsection{Key findings}
\begin{enumerate}
    \item \textbf{Lack of a centralized configuration knowledge base:} There is no coordinated effort to harmonize Kubernetes configuration recommendations across standardization bodies, nor are vendors required to contribute to a shared repository comparable to vulnerability databases such as NVD\footnote{\url{https://nvd.nist.gov/}}, EUVD\footnote{\url{https://euvd.enisa.europa.eu/}}, or OSV\footnote{\url{https://osv.dev/}}. Consequently, multiple hardening guidelines coexist, differing in scope, level of detail, and focus areas (RQ1). Consolidating these heterogeneous recommendations requires substantial manual effort, which is currently delegated to practitioners operating Kubernetes environments.
    \item \textbf{Fragmented detection coverage of configuration scanners:} In the absence of a central source of configuration issues, scanner vendors maintain proprietary rule sets. As a result, different tools detect different subsets of configuration issues, with only partial overlap (RQ2). The choice of scanner therefore directly determines which configuration issues are identified, leading to systematic blind spots when a single tool is used.
    \item \textbf{Non-standardized configuration risk assessment:} Configuration scanners also differ significantly in their risk assessment approaches. Although the Common Configuration Scoring System (CCSS) defines a standardized method for assessing configuration risk, it has neither evolved nor been widely adopted for more than 15 years. The limited adoption of CCSS, including by organizations publishing hardening guidelines (e.g., CIS), has resulted in proprietary, non-comparable risk models across scanners (RQ3). We adopted CCSS for its structured scoring methodology---the systematic decomposition into access vector, complexity, and impact---rather than the currency of its specification. An updated configuration scoring framework remains an open research need.
\end{enumerate}


\subsection{Implications for practitioners}
Practitioners should not rely on a single hardening guideline, as individual guidelines cover only subsets of relevant configuration issues and often provide insufficient guidance on concrete configuration values. Combining multiple guidelines is therefore necessary to improve coverage.

Similarly, scan results obtained from a single static configuration scanner should be treated as incomplete. Coverage can be improved by using multiple scanners or by extending existing tools with custom rules aligned with selected hardening guidelines.

Finally, scanner-provided risk assessments should not be used in isolation. Given the absence of standardized scoring, organizations should perform contextual risk assessments that account for asset criticality, threat models, and potential business impact.

\subsection{Research gaps and future work}
First, configuration issues defined across guidelines and scanners lack usable identifiers, preventing systematic alignment and comparison. Although the Common Configuration Enumeration (CCE) standard provides a suitable identification mechanism, it has not been adopted for Kubernetes configurations. Future work should investigate adoption barriers and alternative identification schemes.

Second, configuration risk scoring remains underdeveloped compared to vulnerability scoring. Future research should evaluate whether existing scoring systems can be adapted to configuration issues or whether new models are required.

Third, structured and reproducible methods for assessing the impact of configuration issues are largely missing. Developing systematic impact assessment methodologies is a prerequisite for reliable configuration risk scoring and prioritization.

\section{Related Work}
Existing literature on Kubernetes security primarily focuses on categorizing configuration issues, analyzing security smells, and evaluating tool performance in limited contexts.

\paragraph{Quantitative and qualitative studies on structuring configuration issues}
Shamim et al.~\cite{Shamim_Bhuiyan_Rahman_2020} identified eleven security practices as a result of a literature review analyzing 104 internet artifacts. The dataset consists of documentation and presentations and mostly blog posts~\cite{dataset_Shamim_Bhuiyan_Rahman_2020}.

In Bose et al.~\cite{Bose_Rahman_Shamim_2021}, the authors performed a qualitative analysis on the frequency of appearance of security defects. They did so by looking at over 5000 commits from 38 open source repositories and determining if a commit was security-related. They do not attempt to identify the actual configuration issues or categorize them. 

Rahman et al.~\cite{Rahman_Shamim_Bose_Pandita_2023} conducted an empirical study analyzing over 2000 Kubernetes manifests mined from 92 open source software repositories to systematically categorize configuration issues in Kubernetes manifests. Additionally, they created a static analysis tool (SLI-KUBE) to quantify the frequency of the identified configuration issues. As a result, they have uncovered eleven categories, which are represented by eleven specific rules in their tool.

Ponce et al.~\cite{Ponce_Soldani_Astudillo_Brogi_2022} conducted a review of 58~scientific and non-scientific literature publications on security smells in microservice-based applications and distilled ten common smells that violate security properties. In the subsequent work by Dell’Immagine et al.~\cite{Dell_Immagine_Soldani_Brogi_2023}, they translated these ten security smells to Kubernetes deployments. For this, they implemented their own tool to analyze these smells both statically and dynamically. They build upon two open-source tools (Checkov and Kubesec.io) for the analysis, which we also evaluated individually against our benchmark. Their work provides a good analysis of smells in microservice-based applications. They do not inspect issues in adjacent areas in a Kubernetes environment, like sensitive RBAC configurations, multi-tenancy, or admission control in general.

A comprehensive study on compliance models, security controls, and threats in a cloud environment was done by Hendre and Joshi~\cite{Hendre_Joshi_2015}. The authors then created an ontology based on these controls and threats, which in turn could be used to recommend cloud security policies. Our work is more focused on technical controls specific to Kubernetes clusters.

An analysis of configuration issues and vulnerabilities and their impact on the runtime security of a container was conducted by Pothula et al.~\cite{Pothula_Kumar_Kumar_2019}. Following the analysis, they propose a mapping of security controls for the runtime security of containers, linking the aforementioned vulnerabilities to a set of controls and best practices. 

Theodoropoulos et al.~\cite{Theodoropoulos_2023} provide a comprehensive survey on cloud security in general. They perform an exploratory analysis that identifies key security aspects in the cloud and map these onto available security features. These, in turn, are mapped to numerous contemporary solutions to implement them. Compared to our work, which focuses on Kubernetes, their identified aspects are focused on the cloud service provider (CSP). Consequently, their solution encompasses technologies not native to Kubernetes, and they do not discuss the specific configuration issues in Kubernetes in detail.

\paragraph{Environments for structured security analysis and evaluation of security tools}

To address the shortage of environments to test security attacks and defenses Minna and Massacci~\cite{Minna_Massacci_2022} created a testbed, which can generate and deploy test cloud environments. Kubernetes is one of the supported technologies. In this work, the authors focus on the generation of susceptible cloud environments in general, where it’s possible to deploy vulnerable Kubernetes components and nodes.

Kamieniarz et al.\cite{Kamieniarz_Mazurczyk_2024} analyzed the security posture of four different Kubernetes cluster deployments using different Kubernetes security tools. Their analysis is focused on identifying security issues in default deployments of Kubernetes comparing deployments using EKS, GKE, RKE2 and K3s. As part of their study they also investigated CIS compliance. However, they stated that their results were inconclusive due to shortcomings of the open-source tools they used for their experiments.

Kapetanidou et al.~\cite{Kapetanidou_Nizamis_Votis_2025} evaluated six Kubernetes security tools, which are also covered in our study. However, their analysis is focused on performance and resource consumption, with only a high-level estimate on the number of detected threats in a single environment. For their research they relied on Kubernetes Goat~\cite{kubernetes-goat}, which is vulnerable by design to educate about Kubernetes security and is not intended for benchmarking a broad spectrum of configuration issues. Our work examines the same tools and additional ones, focusing on the depth and breadth of their security checks and the configuration issues they cover.

\section{Conclusions}
This paper presents a comprehensive study of Kubernetes hardening guidelines and benchmarks, identifying common configuration recommendations and highlighting differences across guidelines from multiple organizations. Based on this analysis, we encode representative configuration issues in Kubernetes manifests and benchmark 10 popular open-source static scanners with respect to coverage, detection accuracy, and risk assessment. Our study addresses a gap in prior work, which has primarily focused on discovering new configuration issues or evaluating scanner performance in isolation.

Our findings reveal a lack of consolidation of Kubernetes configuration issues and the absence of a coordinated approach to risk assessment. As a result, scanners detect different subsets of configuration issues and assign divergent risk scores even for similar or identical findings. This fragmentation directly impacts the effectiveness of configuration scanning tools.

The core implication of this work is that Kubernetes hardening efforts suffer from limited coordination across guidelines and tools. We therefore recommend that practitioners consult multiple hardening guidelines, use more than one scanner, and apply context-aware risk assessments for identified issues. More broadly, our results highlight the need for the research community to develop standardized methods for consolidating and scoring configuration issues that align with the practical demands of modern cybersecurity operations.

%
%
%

\bibliographystyle{splncs04}
\bibliography{literature}

\end{document}